\documentclass[sigconf]{acmart}
\acmConference[ESEC/FSE 2022]{The 30th ACM Joint European Software Engineering Conference and Symposium on the Foundations of Software Engineering}{14 - 18 November, 2022}{Singapore}
\AtBeginDocument{%
  \providecommand\BibTeX{{%
    \normalfont B\kern-0.5em{\scshape i\kern-0.25em b}\kern-0.8em\TeX}}}

\setcopyright{acmcopyright}
\copyrightyear{2018}
\acmYear{2018}
\acmDOI{XXXXXXX.XXXXXXX}

\acmConference[Conference acronym 'XX]{Make sure to enter the correct
  conference title from your rights confirmation emai}{June 03--05,
  2018}{Woodstock, NY}
\acmBooktitle{Woodstock '18: ACM Symposium on Neural Gaze Detection,
 June 03--05, 2018, Woodstock, NY} 
\acmPrice{15.00}
\acmISBN{978-1-4503-XXXX-X/18/06}

\usepackage{listings}
\usepackage{xcolor}
\definecolor{codegreen}{rgb}{0,0.6,0}
\definecolor{codegray}{rgb}{0.5,0.5,0.5}
\definecolor{codepurple}{rgb}{0.58,0,0.82}
\definecolor{backcolour}{rgb}{0.95,0.95,0.92}

\lstdefinestyle{mystyle}{
    backgroundcolor=\color{backcolour},   
    commentstyle=\color{codegreen},
    keywordstyle=\color{magenta},
    numberstyle=\tiny\color{codegray},
    stringstyle=\color{codepurple},
    basicstyle=\ttfamily\footnotesize,
    breakatwhitespace=false,         
    breaklines=true,                 
    captionpos=b,                    
    keepspaces=true,                 
    numbers=left,                    
    numbersep=5pt,                  
    showspaces=false,                
    showstringspaces=false,
    showtabs=false,                  
    tabsize=2
}
\begin{document}

\title{WikiDoMiner: \textbf{Wiki}pedia {Do}main-specific \textbf{Miner}}

\author{Saad Ezzini}
\affiliation{%
  \institution{University of Luxembourg}
    \country{Luxembourg}
}
\email{saad.ezzini@uni.lu}

\author{Sallam Abualhaija}
\affiliation{%
  \institution{University of Luxembourg}
  \country{Luxembourg}
  }
  \email{sallam.abualhaija@uni.lu}

\author{Mehrdad Sabetzadeh}
\affiliation{%
  \institution{University of Ottawa}
  \country{Canada}
}\email{m.sabetzadeh@uottawa.ca}

\renewcommand{\shortauthors}{Ezzini, et al.}


\begin{abstract}

We introduce WikiDoMiner --  a tool for automatically generating domain-specific corpora by crawling Wikipedia. WikiDoMiner helps requirements engineers create an external knowledge resource that is specific to the underlying domain of a given requirements specification (RS). Being able to build such a resource is important since domain-specific datasets are scarce. WikiDoMiner generates a corpus by first extracting a set of domain-specific keywords from a given RS, and then querying Wikipedia for these keywords. The output of WikiDoMiner is a set of Wikipedia articles relevant to the domain of the input RS.
Mining Wikipedia for domain-specific knowledge can be beneficial for multiple requirements engineering tasks, e.g., ambiguity handling, requirements classification, and question answering. 
WikiDoMiner is publicly available on Zenodo under an open-source license ~\cite{WikiDoMiner:22}. 
\end{abstract}

\ccsdesc[500]{Software and its engineering~Requirements analysis}
\ccsdesc[500]{Computing methodologies~Language resources}
\keywords{Requirements Engineering, Natural-language Requirements, Natural Language Processing, Domain-specific Corpus Generation, Wikipedia}

\maketitle

\section{Introduction}\label{sec:introduction}

Requirements specifications (RSs) vary considerably across domains in large part due to the specific terminology associated with each domain~\cite{Ferrari:19,Abualhaija:19}. 
Several requirements engineering (RE) tasks can be performed more accurately when scoped to a specific domain. For example, Winkler and Vogelsang~\cite{Winkler:18} propose an automated solution for classifying requirements and non-requirements for the automotive domain. Ferrari et al.~\cite{Ferrari:18} investigate defects in requirements for the railway domain. Ezzini et al~\cite{Ezzini:21} propose a domain-specific method for handling ambiguity in  requirements.
Addressing RE automation in a domain-specific manner usually necessitates domain-specific knowledge resources. Such resources are nonetheless often unavailable, since domain-specific \hbox{datasets in RE are scarce.} 

In the recent RE literature, there is an increasing reliance on natural language processing (NLP) technologies for automation, leading to the rapidly emerging research area of NLP4RE~\cite{Zhao:21}. Meanwhile, NLP is shifting towards the application of large-scale language models, e.g.,  BERT~\cite{Devlin:18}, for solving downstream NLP tasks such as question answering, natural language inference, and paraphrasing~\cite{Wang:18}. 
Language models are often pre-trained on large bodies of generic text. For instance, the original BERT model is pre-trained on the entire (English) Wikipedia and the BookCorpus. This way, pre-trained language models would learn about word co-occurrences as well as syntactic and semantic regularities in passages. Pre-trained language models can then be fine-tuned for solving downstream tasks. Fine-tuning is the process of exposing a pre-trained model to another dataset that is task-specific and/or in-domain~\cite{Ezzini:22}.  

Due to this evolutionary development in NLP, many NLP4RE solutions -- even some of the most recent ones -- need to be re-examined and revamped to fit the new technological trend. The reason is not only to improve the accuracy of the existing NLP4RE solutions, but also to avoid relying on NLP libraries that will soon be outdated, in turn leading to maintenance headaches and upgrading difficulties. Another essential aspect that is likely to impact the current NLP4RE literature is  reusability. The current implementation tendency in view of the available large-scale language models is towards Python-based libraries. To enable better reusability of the existing NLP4RE solutions, it is advantageous to have a more homogeneous implementation in Python, even when similar libraries are available in other languages, e.g., Java.

In this paper, we present \emph{WikiDoMiner} (\textbf{Wiki}pedia \textbf{Do}main-specific \textbf{Miner}). Given an RS as input, WikiDoMiner automatically generates a domain-specific corpus from Wikepedia, without any a-priori assumptions about the domain of the input RS. WikiDoMiner is a re-implementation of the corpus generator in an earlier research prototype, MAANA~\cite{Ezzini:21}. MAANA is an automated ambiguity handling tool which uses frequency-based heuristics to detect coordination and prepositional-attachment ambiguity. In that context, a large domain-specific corpus is needed for estimating word frequencies. In our ongoing research since MAANA, we have increasingly needed domain-specific corpus generation, not for frequency-based statistics but rather for fine-tuning pre-trained language models. This prompted us to build and release WikiDoMiner as a stand-alone tool and a more robust and usable alternative to the corpus generator in MAANA. MAANA's corpus generator is Java-based. Furthermore, it requires a local dump of Wikipedia installed as an SQL database. This consumes significant resources and makes both the installation and (re-)use of MAANA complex. WikiDoMiner lifts this major limitation and further, by virtue of being Python-based, is much easier to use alongside language models.

\begin{figure*}
\begin{center}
\includegraphics[width=.99\textwidth]{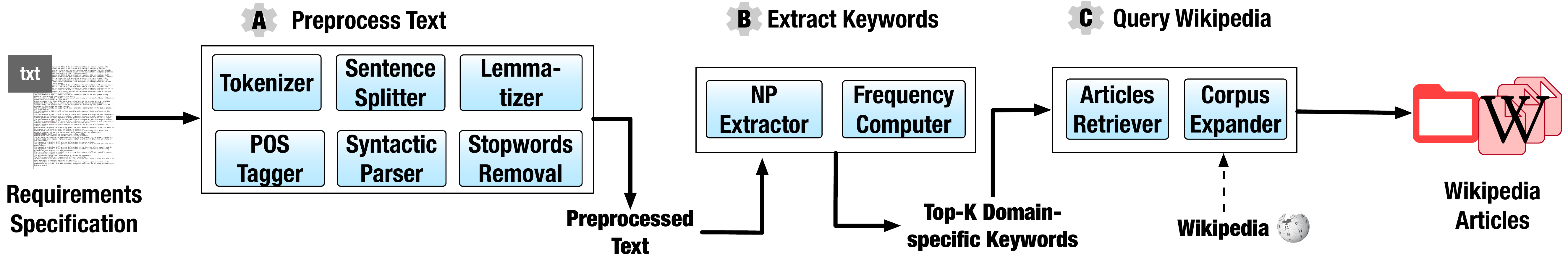} 
\end{center}
\caption{Tool Architecture.}
\label{fig:tool} 
\end{figure*}

In the rest of this  paper, we outline the workings of WikiDoMiner and demonstrate the tool's application in two different domains.

\section{Tool Architecture}\label{sec:architecture}

WikiDoMiner is a usable prototype tool for generating domain-specific corpora. 
Figure~\ref{fig:tool} shows an overview of WikiDoMiner architecture. 
The tool is implemented in Python 3.7.13~\cite{Rossum:09} using Google Colab\footnote{\url{https://colab.research.google.com/?utm_source=scs-index}}.
Below, we discuss the different steps of the tool marked A -- C in Figure~\ref{fig:tool}. 

\subsection{Preprocess Text}
In the first step, we parse the textual content of the input RS and preprocess the text. To do so, we apply an NLP pipeline composed of six modules, four of which are related to parsing and normalizing the text, and two are for performing syntactic parsing. These modules include: A \textit{tokenizer} splits the text into different tokens (e.g., commas and words), \textit{sentence spitter} identifies the boundaries of sentences in the running text (e.g., a sentence in English can end with a period), \textit{lemmatizer} finds the canonical form of a word (e.g., the singular word ``communication'' is the canonical form of its plural variant ``communications'' and the infinitive ``transmit'' is the canonical form for its past-tense variant ``transmitted''), and finally, a \textit{stopwords removal} module removes the stopwords such as articles (``the'') and prepositions (e.g., ``in''). To perform syntactic analysis, we further apply: \textit{POS tagger} that assigns a part-of-speech tag for each token (e.g., the tag VBD is assigned to ``transmitted'' indicating a past-tense verb), and a \textit{syntactic parser} that identifies the syntactic units in the text (e.g., ``the notification service'' is a noun phrase -- NP).

To operationalize the NLP pipeline, we use the Tokenizer, Porter Stemmer and WordNet Lemmatizer available in NLTK 3.2.5~\cite{NLTK}. We further apply Python RE 2.2.1 regex library\footnote{\url{https://docs.python.org/3/library/re.html}}, in addition to available modules in SpaCy 3.3.0~\cite{spacy} including the English stopwords list, Tokenizer, NP Chunker, Dependency Parser, and Entity Recognizer.

\subsection{Extract Keywords} 
In this step, we extract a set of keywords that are representative for the underlying domain. To do that, we adapt a glossary extraction method from the RE literature~\cite{Arora:17}. The basic idea in this step is to collect the noun phrases in the input RS, and sort them according to their frequency of use. To ensure that these keywords are domain-specific, WikiDoMiner applies two measures. First, we remove from the list any keyword that is available in WordNet~\cite{Miller:95,Fellbaum:98}, which is a generic lexical database for English. The intuition of this step is to remove very common words that are not representative of the underlying domain. For instance, the word ``rover'' exists in WordNet\footnote{\url{http://wordnetweb.princeton.edu/perl/webwn?s=rover&sub=Search+WordNet&o2=&o0=1&o8=1&o1=1&o7=&o5=&o9=&o6=&o3=&o4=&h=}} as a noun referring to ``someone who leads a wandering unsettled life'' or ``an adult member of the Boy Scouts movement''. These two meanings do not fit the ``rover'' in the ``lunar rover'' context, and the NP ``lunar rover''. This way, we filter out the word ``rover'' when it occurs alone (i.e., ``the rover''), and keep it as part of the NP (``lunar rover''). We note that the the NP ``lunar rover'' is not available in WordNet, but is in Wikipedia\footnote{\url{https://en.wikipedia.org/wiki/Lunar_rover}}. 

As a second measure, WikiDoMiner computes term frequency/inverse document frequency (TF/IDF)~\cite{McGill:83} instead of mere frequency. TF/IDF is a traditional method that is often applied in the context of information retrieval (IR) to assign a score reflecting the importance of words to a specific document in a document collection. In WikiDoMiner, the importance of the words (NPs in our case) indicates that the words are significant for the underlying domain. We note that IDF is computed only when there are multiple documents from other domains available. Otherwise the TF/IDF scores are similar to term frequencies. 
Once the TF/IDF scores are computed, we sort the keywords in descending order of these scores and select the top-$K$ keywords. While the default value applied by WikiDoMiner is $K=50$, we show in the demo of the tool that this parameter can be configured by the user according to the intended application. 

We implement the different modules using WordNet from NLTK 3.2.5~\cite{NLTK}, 
and TF-IDF transformation from Scikit-learn 1.0.2~\cite{scikit-learn}.

\subsection{Query Wikipedia}
In this step, we use the keywords from the previous set to query Wikipedia and collect the relevant articles which will then constitute our final domain-specific corpus. 

\begin{figure}[!t] 
\begin{center}
\includegraphics[width=.48\textwidth]{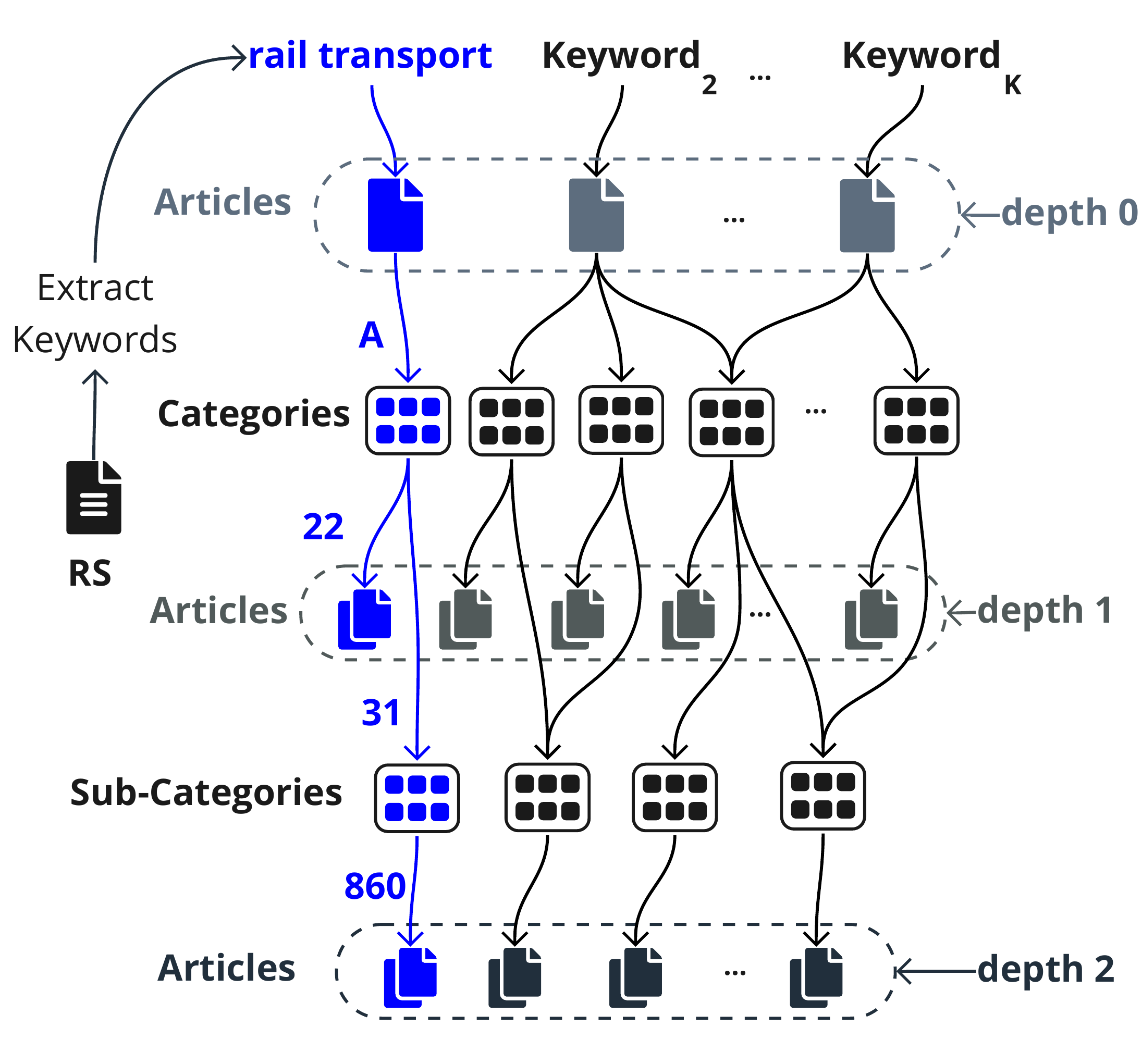} 
\end{center}
\caption{Illustration of Traversing Wikipedia Categories (Example Keyword: ``rail transport'').}
\label{fig:wiki} 
\end{figure}

To better understand this step, we first explain the structure of a category in Wikipedia, illustrated in Figure~\ref{fig:wiki}. Each article in Wikipedia belongs to one or more categories. Each category contains a set of articles and sub-categories. 
To illustrate, assume that we are querying Wikipedia for the keyword ``rail transport'' within the ``Railway'' domain. 
Our first hit will be a page titled ``Rail Transport''\footnote{\url{https://en.wikipedia.org/wiki/Rail_transport}}. Note that we refer to a page in Wikipedia as an \textit{article}. If we view the category structure for this article\footnote{\url{https://en.wikipedia.org/wiki/Category:Rail_transport}}, we find out that it belongs to a category under the same name ``Rail Transport'', i.e., Category A in Figure~\ref{fig:wiki}. Inside this category, there are 31 sub-categories such as ``Locomotives'', ``Rail Infrastructure'', and ``Electric rail transport''.  Category A contains 22 other pages alongside the above mentioned pages, such as ``Bi-directional vehicle'' and ``Pocket wagon''. Viewing the structure of a sub-category, e.g., ``Rail Infrastructure'' will show us again the available pages and sub-categories.

In WikiDoMiner, the result of querying Wikipedia for a given keyword is a Wikipedia article whose title partially matches the keyword. We consider the title of a Wikipedia article as partially matching the keyword if they have some overlap. 
For example, if we query Wikipedia for the keyword ``Efficiency of rail transport'', then we will retrieve the same article mentioned above whose title, ``Rail Transport'', partially matches the keyword. 

For each keyword, we retrieve from Wikipedia a matching article if applicable. Some applications might require that the domain-specific corpus be sufficiently large. For example, to accurately estimate the frequencies of words co-occurrences, one needs a large corpus~\cite{Jurafsky:20}. Similarly, to pre-train a domain-specific language model, a large text body should be available. Therefore, we expand our corpus by defining a configurable parameter \textit{depth} to control the level of expansion, thus allowing the user to adjust the size and relevance of the corpus based on their needs. The minimal depth $\textit{depth} =0$ can be used to extract directly matching articles only (leading most often to a few hundred articles). 
WikiDoMiner further retrieves, for each matching article, all articles in the same categories for $\textit{depth} = 1$ (e.g., the two other pages in the example above), subcategories of $\textit{depth} = 2$, sub-subcategories of $\textit{depth} = 3$, and so on.

Specific details of our implementation are as follows. We use the Wikipedia 1.4.0\footnote{\url{https://wikipedia.readthedocs.io/}} and Wikipedia-API 0.5.4\footnote{\url{https://wikipedia-api.readthedocs.io/}} libraries to query Wikipedia. Other libraries which we use but which are not necessary to run the tool include PyPDF2 2.2.0\footnote{\url{https://pypdf2.readthedocs.io/}} to read requirements documents in PDF format, the word2vec similarity feature in SpaCy 3.3.0 library~\cite{spacy}, and the WordCloud 1.5.0\footnote{\url{https://amueller.github.io/word_cloud/}} library to visualize the most prevalent words in the extracted corpora.

\section{Application} \label{sec:application}

In this section, we apply WikiDoMiner to automatically generate domain-specific corpora for two distinct domains, namely, railway and transportation. 
We further assess how representative the corpus generated for each of these domains is. We do so by computing the semantic relatedness of each domain-specific corpus against RSs from the same domain other than those used for generating the corpus. Generating a domain-specific corpus is not a frequent activity. In practice, requirements engineers would typically have a small set of RSs from a given domain at the time of generating a domain-specific corpus and would utilize this corpus to perform activities on other RSs not involved in the generation process. 

\subsection{Data Collection} \label{sec:datacollection} 
For the two domains considered in this section, we collected a total of six RSs from the PURE dataset~\cite{ferrari:17}, with three RSs from each domain. One RS is used for generating the corpus and the others are used for evaluating  semantic relatedness against the resulting corpus.

In the following we list the six RSs: 
\begin{itemize}
    \item From the railway domain, we used RS1 (\textit{ERTMS}) about \textit{train control}, RS2 (\textit{EIRENE SYS 15}) and RS3 (\textit{EIRENE FUN 7}) both about \textit{digital radio standard for railway}. 
    \item From the transportation domain, we used RS4 (\textit{CTC NETWORK}) about \textit{traffic management infrastructure}, RS5 (\textit{PONTIS}) about \textit{highway bridge information management}, and RS6 (\textit{MDOT}) about \textit{transportation info management}.
\end{itemize}

\begin{figure*}
\begin{center}
\includegraphics[width=.97\textwidth]{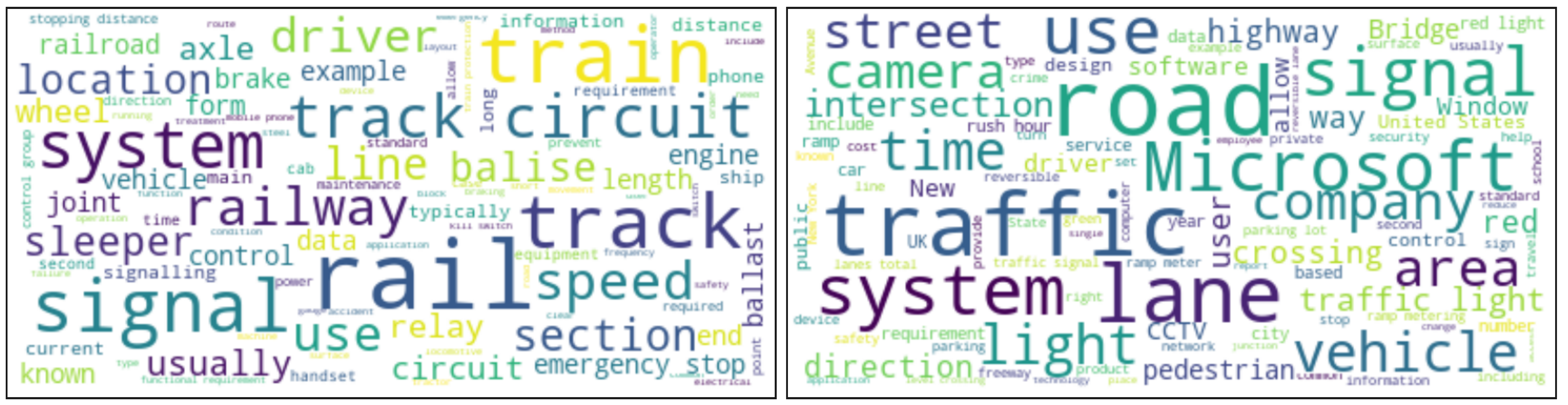} 
\end{center}
\caption{Word-cloud Visualization of Domain-specific Corpora (Left-hand Side -- Railway Domain, and Right-hand Side -- Transportation Domain).}
\label{fig:corpora} 
\end{figure*}

\subsection{Domain-specific Corpora}
For illustration, we centre our discussion around the railway domain. We generate the corpus from RS1, and evaluate the relatedness on RS2 and RS3. The first step in WikiDoMiner is to preprocess RS1.
WikiDoMiner then extracts a set of keywords based on their TF/IDF scores. Examples of such keywords include \textit{trainborne equipment} and \textit{emergency brake}. We select the top-$K$ keywords, where $K=50$. 

The next step is to query the keywords on Wikipedia. For our set of keywords in this domain, we retrieve 15 matching articles. We then set the configuration parameter \textit{depth} to 1. Following this, we collect for each article that matches a keyword the articles in the respective categories (see Figure~\ref{fig:wiki}). 
Finally, we collected a total of \textbf{686 articles}, which are considered as our domain-specific corpus.  

We apply WikiDoMiner on RS4 (from the transportation domain) in a similar manner. The two resulting corpora are depicted in Figure~\ref{fig:corpora} as word clouds. We show for each domain the main terms that frequently occur in the corpus. We see that the keywords \textit{rail}, \textit{track}, \textit{train}, \textit{railway}, and \textit{railroad} characterize the railway corpus, while the transportation corpus is characterized by the keywords \textit{traffic}, \textit{road}, \textit{street}, and \textit{lane}. We note that the railway domain can be regarded  as a sub-domain of the transportation domain. This observation is highlighted through the frequent terms that the two corpora have in common in Figure~\ref{fig:corpora}, such as \textit{signal}, \textit{system}, \textit{vehicle}, and \textit{driver}.

\subsection{Domain Representativeness}
To examine how representative the resulting domain-specific corpora are, we compute semantic relatedness as follows. 
We first transform each article in the corpus into a vector representation using word2vec. We do the same for the test RS. Then, we compute the cosine similarity between the vector representing the (test) RS and the vector representing each article. 
In the following, we report the minimum, average, and maximum cosine similarity scores for each domain:

\begin{itemize}
    \item Railway domain (cosine similarity between the railway corpus and test RSs): min=0.27, \textbf{average=0.94}, and max=0.98
    \item Transportation domain (cosine similarity between the transportation corpus and test RSs): min=0.67, \textbf{average=0.95}, and max=0.99.
\end{itemize}

Our results show that the domain-specific corpora are, on average, highly similar to the test (unseen) RSs not used for generating the corpora. 
In particular, the average semantic similarity is $\geq 0.94$, indicating that many articles in the corpus are relevant to the test RSs. The minimum score of 0.27 in the railway domain implies that there are articles in the corpus which are more document-specific, i.e., more relevant to the RS that induced the corpus but having little in common with the test RSs. Note that, despite some document-specific articles being present in the generated corpus, the very high average semantic similarity  ($\geq 0.94$) indicates that such articles are a small minority and thus do not have a significant negative impact on the in-domain usability of the generated corpus.

The gap seen  between the minimum scores reported for the two domain-specific corpora can be explained by the following: As mentioned in Section~\ref{sec:datacollection}, all RSs from the transportation domain in our collection  are on the topic of traffic and transportation information management. This leads to extracting many keywords related to information management. In contrast, the RSs in our collection from the railway domain are tailored to more specific topics, namely train control and digital radio standard for railway. This in turn leads to extracting document-specific terms which are related to train control (i.e., the topic of the RS used for corpus generation) but not so much to digital radio standard for railway (i.e., the topic of the test RSs). 
To summarize, our experiments show that WikiDoMiner has successfully generated representative \hbox{corpora for two distinct domains.}

\section{Conclusion}

We presented WikiDoMiner, a tool for automatically generating  domain-specific corpora from Wikipedia. Our current implementation is a significantly enhanced and usable adaptation of the corpus generation component briefly outlined in our earlier work~\cite{Ezzini:21}. WikiDoMiner extracts keywords from a given requirements specification (RS) and then queries these keywords in Wikipedia. For each keyword, WikiDoMiner looks for a matching article whose title has some overlap with the keyword. To expand the corpus, we provide the user with the possibility to configure a parameter \textit{depth} that controls how deeply the Wikipedia category structure should be traversed. We assess the relatedness of the resulting corpora to RSs different from those used for corpus generation. Our empirical results show that, across two distinct domains, WikiDoMiner yields an average semantic relatedness of $\geq 0.94$ for in-domain analysis.  

In the future, we plan to utilize WikiDoMiner for addressing new analytical problems beyond ambiguity analysis. Notable target problems include 
question answering \hbox{and requirements classification.}

\begin{acks}
This work was funded by Luxembourg's National Research Fund (FNR) under the grant BRIDGES18/IS/12632261 and NSERC of Canada under the Discovery and Discovery Accelerator programs. We are grateful to the research and development team at QRA Corp. for valuable insights and assistance.
\end{acks}

\newpage
\balance
\bibliographystyle{ACM-Reference-Format}
\bibliography{main}

\end{document}